\documentclass[aps,prd,twocolumn,superscriptaddress]{revtex4-2}
\usepackage{graphicx}
\usepackage{dcolumn}
\usepackage{bm}

\usepackage{hyperref,url}
\usepackage{comment}
\usepackage{lipsum}
\usepackage{amsmath}
\usepackage{color}
\usepackage[caption=false]{subfig}
\usepackage{pifont}
\begin{document}

\title{DC response of an interferometer topology with an L-shaped cavity: a tabletop study}
\author{Junlang Li}\thanks{These authors contributed equally to this work.}
\affiliation{School of Physics and Technology, Wuhan University, Wuhan 430072, China}
\affiliation{School of Physics and Astronomy, Beijing Normal University, Beijing, 100875, China}
\author{Jiehong Huang}\thanks{These authors contributed equally to this work.}%
\author{Xinyao Guo}
\author{Haixing Miao}
\affiliation{Frontier Science Center for Quantum Information, Department of Physics, Tsinghua University, Beijing, 100084, China}
\author{Yuchao Chen}
\author{Xiaoman Huang}
\author{Yuan Pan}
\author{Chenjie Zhou}
\affiliation{School of Physics and Astronomy, Beijing Normal University, Beijing, 100875, China}
\author{Raffaele Flaminio}
\affiliation{Laboratoire d'Annecy de Physique des Particules,  CNRS/IN2P3
9 Chemin de Bellevue - BP110
74941, Annecy-le-Vieux, Annecy, France}
\author{Jameson Graef Rollins}
\affiliation{LIGO Laboratory, California Institute of Technology, Pasadena, California 91125, USA}
\author{Bram Slagmolen}
\affiliation{OzGrav, Australian National University, Canberra, Australian Capital Territory 0200, Australia}
\author{Fan Zhang}
\email{fnzhang@bnu.edu.cn}
\affiliation{Institute for Frontiers in Astronomy and Astrophysics, Beijing Normal University,
Beijing 102206, China}
\affiliation{School of Physics and Astronomy, Beijing Normal University, Beijing, 100875, China}
\author{Teng Zhang}
\email{tzhang@star.sr.bham.ac.uk}
\affiliation{School of Physcis and Astronomy, University of Birmingham, Birmingham, B15 2TT, United Kingdom}
\author{Mengyao Wang}
\email{mengyao.wang@bnu.edu.cn}
\affiliation{School of Physics and Astronomy, Beijing Normal University, Beijing, 100875, China}



\date{\today}

\begin{abstract}

A new interferometer topology for kilohertz gravitational-wave detection was recently proposed in [Zhang {\it et al.} Phys. Rev. X 13, 021019 (2023)]. The design is based on an L-shaped optical cavity pumped through a Sagnac-like vortex. We report a tabletop experiment that characterizes the interferometer’s optical response near DC. When the laser frequency is locked to the resonance of the L-shaped cavity, we observe that the cavity input coupler becomes effectively transparent, yielding a simple Michelson-like response. Moreover, the Sagnac vortex separates into upper and lower paths, which behave as two independent pumping paths driving the cavity. These observations are in agreement with theoretical predictions. Our results provide an intuitive physical picture of this interferometer topology and offer insight into its lock acquisition strategy.
\end{abstract}

\maketitle


\section{Introduction}

Gravitational-wave (GW) astronomy has progressed rapidly since the first detections by ground-based detectors\,\cite{PhysRevLett.116.061102,PhysRevLett.119.161101,Aasi_2015,GWTC-3, GWTC-4}, offering unprecedented insights into compact-object mergers and the physics of dense matter. In particular, the joint electromagnetic and GW observation of GW170817 event inaugurated the era of multi-messenger astronomy. Due to limited sensitivity of current detectors in the kilohertz regime, the (post-)merger GW signal from colliding neutron stars was not confidently confirmed because the detector noise is dominated at high frequencies by quantum noise originated from the quantum fluctuation of light\,\cite{Chen_2013,RanaRMP, DanilishinLivingReview}. Several quantum-enhanced schemes have been proposed to modify the input-output optics of the Michelson configuration to improve kHz sensitivity\,\cite{Miao18,Martynov19,NEMO2020, O5upgrade}. However, in the conventional Fabry-Perot Michelson configuration, the high-frequency signals are averaged out by the arm cavity response, and even small optical loss in the signal-recycling cavity will introduce additional vacuum noise that severely contaminates the high-frequency signals\,\cite{Miao19}. This imposes a quantum-loss limit of sensitivity at kHz frequency, and cannot be mitigated by simply increasing the arm length due to decreases of the arm cavity bandwidth. 

\begin{figure}[h!]
        \centering    \includegraphics[width=0.8
    \linewidth]{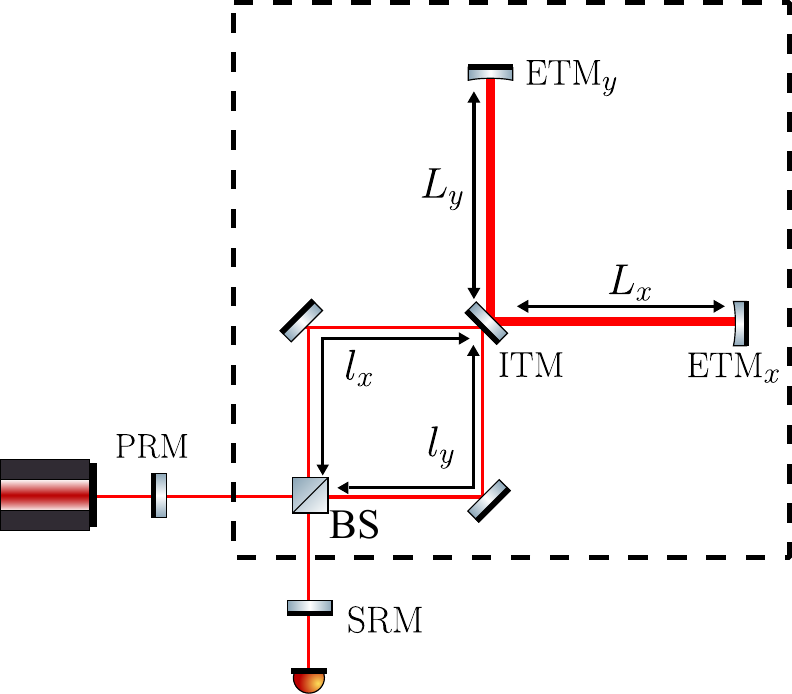}     
        \caption{Layout of the new topology with an L-shaped cavity and Sagnac-like vortex. BS:beam splitter; PRM: power recycling mirror; SRM: signal recycling mirror;  ITM: input test mass; ETM: end test mass. The part enclosed by the dashed black box is the focus of the experimental study.}
        \label{fig:layout1}
\end{figure}

A recently proposed alternative solution is to replace the linear arm cavities with a folded L-shaped optical cavity\,\cite{PhysRevX.13.021019}, similar to the Fox-Smith and synchronous interferometer\,\cite{Smith, Synchronous, Vinet}. By matching the temporal and spatial variation of the GW signal, this topology directly amplifies the kilohertz response at the free spectral range, and becomes immune to the loss in the signal recycling cavity. Its defining feature is an L-shaped cavity pumped through a Sagnac-like vortex, as depicted in Fig.\,\ref{fig:layout1}. This topology requires a sensing and control scheme distinct from that used in current Michelson configuration. A preliminary attempt has been reported in Ref.\,\cite{Guo_2023}, which combines the frontal modulation and an auxiliary control field injected from the dark port to obtain control signals for the core degrees of freedom. 
A critical step for implementing such a scheme is the development of a robust lock-acquisition procedure\,\cite{Evansthesis, Martynovthesis, Staley2014}, which in turn requires both theoretical modeling and an intuitive understanding of the optical response of the core degrees of freedom. 

Idealized theoretical analysis predicts that, when the end mirrors of the L-shaped cavity are perfectly reflective, the Sagnac-vortex behaves like that of a simple Michelson interferometer; if the laser frequency is locked to the resonance of the cavity, scanning the differential arm length will lead to a Michelson-like optical response.  
To validate this prediction, we perform a tabletop experiment to study the DC response of the interferometer topology, focusing on the common and differential modes within the region enclosed by the dashed box in Fig.\,\ref{fig:layout1}. Our results confirm this simplified theoretical picture and reveal additional features relevant for  control and  practical implementation. 

This paper goes as follows. Section\,\ref{sec:theo} presents the theoretical analysis of the new topology DC optical response. Section\,\ref{sec:experi} describes the experimental setup and compares the measured results with the model. We conclude and discuss broader implications in Sec.\,\ref{sec:summary}.



\section{Theoretical description}\label{sec:theo}
\begin{figure}[h!]
  \subfloat[]{
        \centering \includegraphics[width=0.9\linewidth]{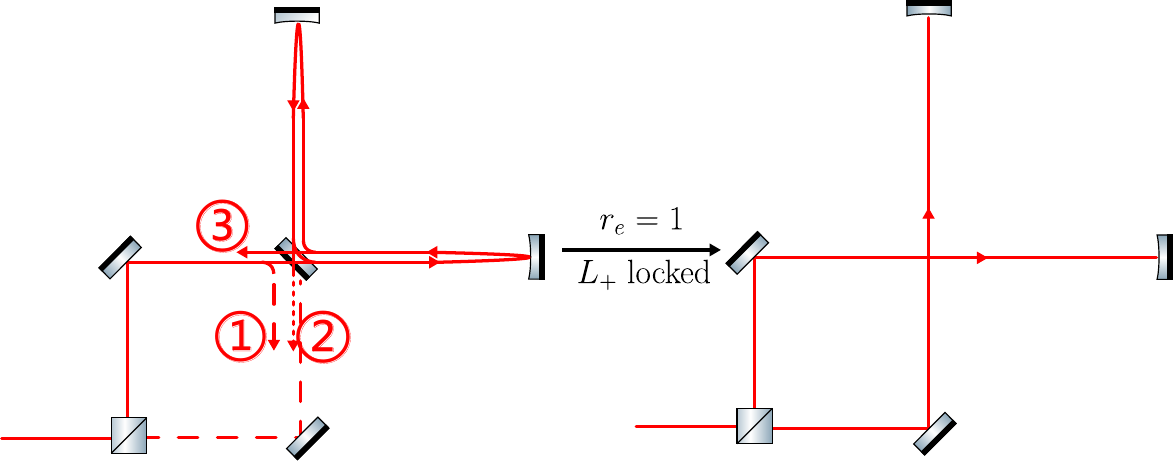}     \label{fig:layout2}
        }
    
    \vspace{0.4em}
    
  \subfloat[]{
        \centering \includegraphics[width=0.9\linewidth]{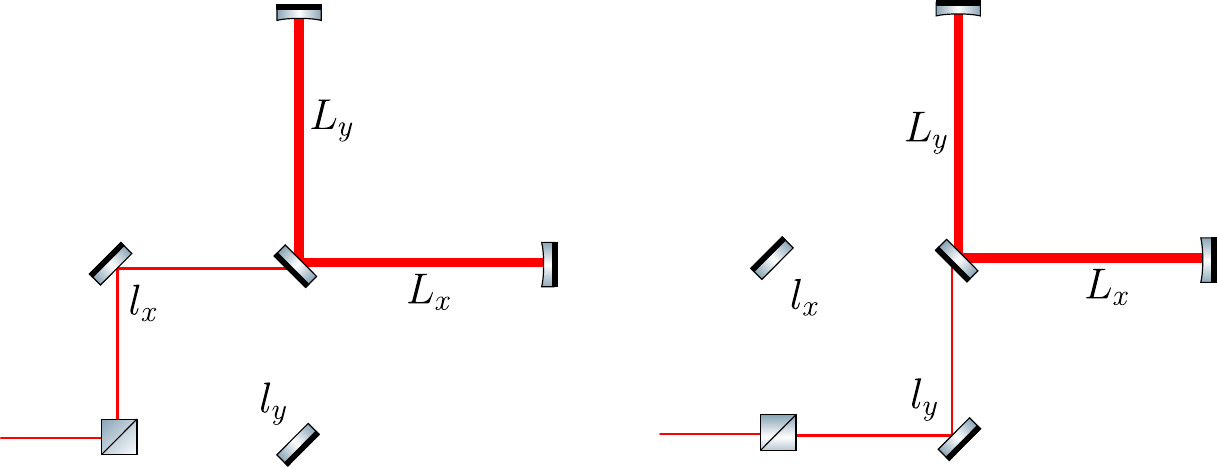}    \label{fig:layout3}
         }
    \caption{(a) Schematic showing that the interferometer with an L-shaped cavity can be regarded as equivalent to a Michelson interferometer. The left panel of the figure only illustrates the circulation of the clockwise incident beam within the L-shaped cavity. The counterclockwise incident beam, indicated by the dashed line, has a symmetric circulation inside the cavity that is not shown. (b) Schematic showing that the input laser can be equivalent to two in-phase beams injected into the interferometer, traveling along the clockwise and counterclockwise paths of the Sagnac vortex, separately, and subsequently pump the L-shaped cavity from opposite directions.
        }

    \label{fig:layoutall}
\end{figure}

In this section, we provide a more detailed description of the interferometer topology and derive the theoretical prediction of its DC optical response. 
As illustrated in Fig.~\ref{fig:layout1}, the input laser beam is split by the beamsplitter into two paths that circulate in opposite directions  within the Sagnac vortex. We denote the clockwise path by $x$, with length of $l_x$ and the counterclockwise path by $y$, with length $l_y$. The two arms of the L-shaped cavity have lengths $L_x$ and $L_y$, respectively. Within this configuration, four longitudinal degrees of freedom can be identified: the Sagnac common mode ($l_+ = l_x + l_y$), the Sagnac differential mode ($l_- = l_x - l_y$), the L-shaped cavity common mode ($L_+ = L_x + L_y$), and the L-shaped cavity differential mode ($L_- = L_x - L_y$). As depicted in the left panel of Fig.\,\ref{fig:layout2}, the laser beam entering the L-shaped cavity for the $x$ path (similarly, for the $y$ path) is reflected in three ways, denoted as \ding{172}, \ding{173}, and \ding{174}. Specifically, \ding{172} corresponds to the prompt reflection by the L-shaped cavity without entering it, whereas \ding{173} and \ding{174} represent the portions of field that enter the cavity and return from different directions.

We first analyze the ideal case in which both ETMs have perfect reflectivity ($r_e = 1$). When the cavity common mode $L_+$ is locked to resonance, the effective reflection coefficients of the L-shaped cavity for the three parts of each incident field can be derived as (see the Appendix for more details):
\begin{align}
&r_{xy} = r_{yx} = 0,
\label{eq:rxy and ryx when re=1} \\
&r_{xx} = e^{i\frac{2\omega_0 L_x }{c}}, \quad
r_{yy} = e^{i\frac{2\omega_0 L_y }{c}},
\label{eq:rxx and ryy when re=1}
\end{align}
where the first subscript labels the input path and the second labels the output path. 
Equation\,\eqref{eq:rxy and ryx when re=1} implies that the prompt reflection \ding{172} cancels exactly with \ding{173}.
\begin{figure*}[t!]
    \centering     
    \subfloat[]{
        \centering 
        \includegraphics[width=0.45\linewidth]{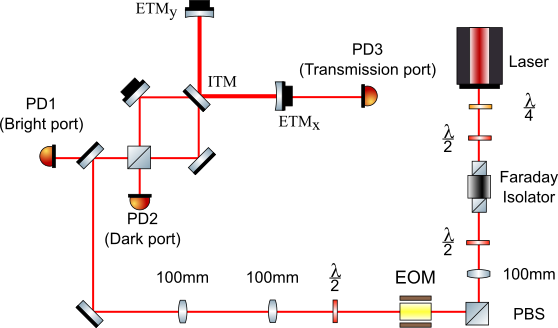}   \label{fig:photo1}
        }
    \subfloat[]{
        \centering
        \includegraphics[width=0.45\linewidth]{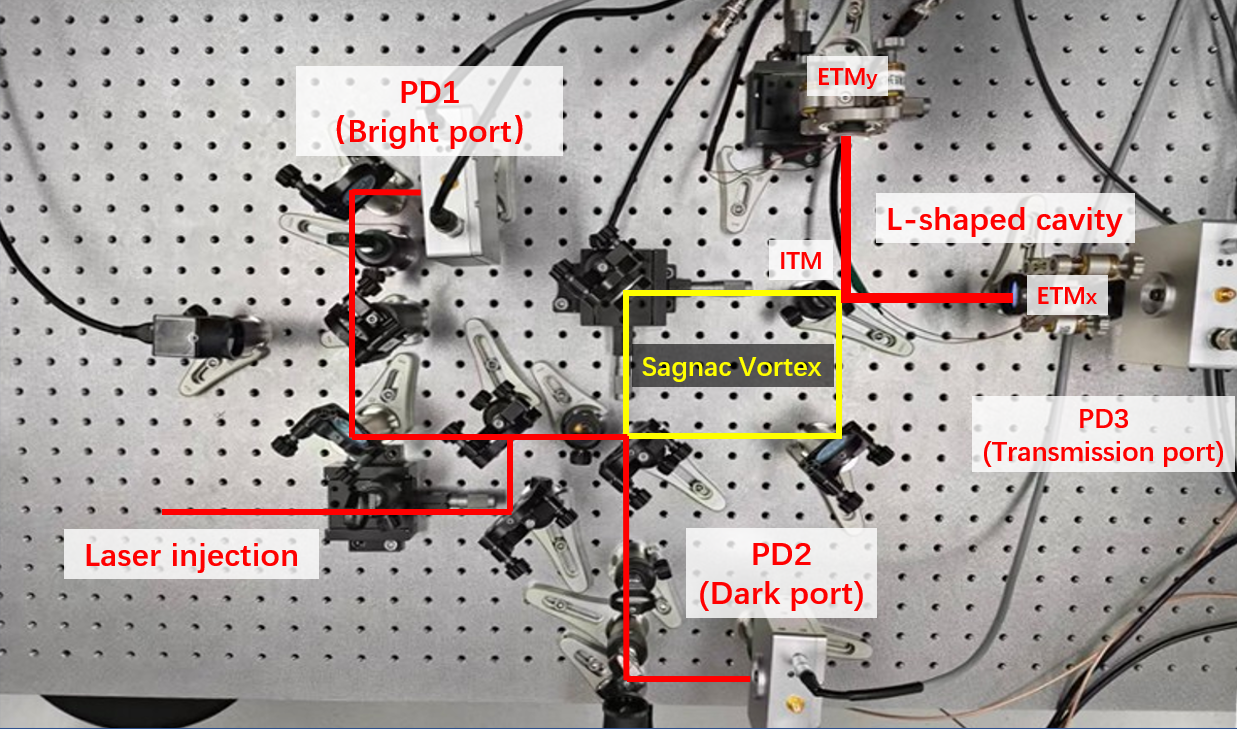}       \label{fig:photo2}
        }
    \caption{Optical layout (a) and setup (b) of the experiment. The L-shaped interferometer contains two main parts, the Sagnac vortex and the L-shaped optical cavity. PBS: polarization beam splitter; EOM: electro-optic modulator; PD: photodetector. }
    \label{fig:layout}
\end{figure*}
From Eqs.\,\eqref{eq:rxy and ryx when re=1} and \eqref{eq:rxx and ryy when re=1}, each beam entering the L-shaped cavity is therefore completely reflected back along its incident direction. Consequently, when viewed from outside the cavity, the behavior of the beams that enter the L-shaped cavity and return is equivalent to that of beams propagating through the lengths $L_x$ and $L_y$ and being directly reflected by a mirror, since in both cases the reflected fields possess identical amplitudes and phases. The entire interferometer thus behaves effectively as a folded Michelson interferometer, as illustrated in the right panel of Fig.~\ref{fig:layout2}. The corresponding DC powers at the dark and bright ports are then given by
\begin{equation}
\frac{S_{\mathrm{dark}}}{S_{\mathrm{in}}} =
\sin^2{\!\left( \frac{\omega_0 \mathcal{L}_-}{c} \right)}, \quad
\frac{S_{\mathrm{bright}}}{S_{\mathrm{in}}} =
\cos^2{\!\left( \frac{\omega_0 \mathcal{L}_-}{c} \right)},
\label{eq:darkDC and brightDC when re=1}
\end{equation}
where $\mathcal{L}_- = L_- + l_-$. Only the sum of $L_-$ and $l_-$ appears and it indicates the degeneracy of the two differential degrees of freedom. This interesting feature was also identified in Ref.\,\cite{Guo_2023} when designing the sensing and control scheme. 

Equation\,\eqref{eq:rxy and ryx when re=1} also indicates that
the Sagnac vortex effectively turns into two independent pumping paths as shown in Fig.~\ref{fig:layout3}. The
analytical expression for the intra-cavity power clearly reflects this feature: 
\begin{equation}
\frac{S_{\mathrm{cavity}}}{S_{\mathrm{in}}} =
\frac{1}{2t_i^2} \left| 1 + r_i e^{ i\frac{\omega_0 \mathcal{L}_-}{c}} \right|^2,
\label{eq:intra-cavityDC when re=1}
\end{equation}
where $r_i$ and $t_i$ denote the reflectivity and transmissivity of the ITM. This expression shows how the intra-cavity power varies as the two paths interfere with each other according to the differential phase difference. Comparing Eq.\,\eqref{eq:intra-cavityDC when re=1} with Eq.\,\eqref{eq:darkDC and brightDC when re=1}, we see that the period of the dark- and bright-port DC signals with respect to $\mathcal{L}_-$ is twice that of the intra-cavity power. This can be understood intuitively as follows. Laser power within arm cavity is coherently pumped by two beams, as illustrated in Fig.~\ref{fig:layout3}. When the common mode is locked, each beam is individually resonant in the L-shaped cavity, and the dependence of the circulating power on ${\cal L}_-$  originates from the interference between the two beams. Regardless of the choice of phase reference within the cavity, the relative phase difference between the beams is determined by the single-trip phase difference accumulated along the clockwise and counterclockwise paths, which is equal to half of the round-trip phase difference between the two paths that could be observed at the dark port.


The above discussion assumes $r_e = 1$. In practice, ETMs have finite transmission ($r_e < 1$). In this case, even with $L_+$ locked, parts \ding{172} and \ding{173} no longer cancel with each other perfectly. Equation\,\eqref{eq:rxy and ryx when re=1} is therefore modified to
\begin{equation}
r_{xy} = r_{yx} = -r_i + \frac{t_i^2 r_i r_e^2}{1 - r_i^2 r_e^2} \neq 0,
\label{eq:rxy and ryx when re<1}
\end{equation}
where the first term corresponds to \ding{172} and the second to \ding{173}. The full interferometer response can thus be separated into two modes: a Michelson mode, defined by $r_{xx}$ and $r_{yy}$, and a Sagnac mode defined by $r_{xy}$ and $r_{yx}$, which behave like the beams in a Sagnac interferometer. At the dark port, the Sagnac-mode reflections from the two paths cancel perfectly, so the DC signal remains a sine function. At the bright port, however, the Sagnac mode contributes a constant offset because it is insensitive to variations in $\mathcal{L}_-$. As a result, the DC signal at the bright port exhibits a sinusoidal variation with a period of $\lambda/2$, while the amplitudes of the peaks are modulated by an additional cosine term with a period of $\lambda$. This picture is indeed followed from the more rigorous analysis, which modifies Eqs.\,\eqref{eq:darkDC and brightDC when re=1} and \eqref{eq:intra-cavityDC when re=1} into 
\begin{equation}
\frac{S_{\mathrm{dark}}}{S_{\mathrm{in}}} =
\frac{t_i^4 r_e^2}{(1 - r_i^2 r_e^2)^2}
\sin^2{\!\left( \frac{\omega_0 \mathcal{L}_-}{c} \right)},
\label{eq:darkDC when re<1}
\end{equation}

\begin{equation}
\frac{S_{\mathrm{bright}}}{S_{\mathrm{in}}} =
\left| -\frac{t_i^2 r_e}{1 - r_i^2 r_e^2}
\cos{\!\left( \frac{\omega_0 \mathcal{L}_-}{c} \right)} +
\frac{r_i - r_i r_e^2}{1 - r_i^2 r_e^2} \right|^2,
\label{eq:brightDC when re<1}
\end{equation}

\begin{equation}
\frac{S_{\mathrm{cavity}}}{S_{\mathrm{in}}} =
\frac{t_i^2}{2(1 - r_i^2 r_e^2)^2}
\left( 1 + r_i^2 r_e^2 +
2r_i r_e \cos{\!\left( \frac{\omega_0 \mathcal{L}_-}{c} \right)} \right).
\label{eq:intra-cavityDC when re<1}
\end{equation}

\section{The experimental study}\label{sec:experi}
In this section, we describe the tabletop experimental setup and compare the measured results with the theoretical predictions presented above.

The optical configuration of the setup is depicted schematically in Fig.~\ref{fig:photo1}, and the physical layout of the core interferometer is shown in Fig.~\ref{fig:photo2}. 
Laser first passes through a Faraday isolator to prevent back-reflections. It then traverses a half-wave plate and a polarization beam splitter, which together regulate the injection power. The beam is subsequently phase modulated by an electro-optic modulator\,(EOM) to generate the control sidebands required for the Pound–Drever–Hall locking scheme\,\cite{Black2001AnIT}. A third half-wave plate rotates the polarization from s- to p-polarization to match the preferred polarization of the L-shaped cavity. The beam is then mode-matched to the optical cavity using two 100\,mm focal-length lenses. Two steering mirrors direct the beam into the interferometer; one of them is a 90\% reflective beam splitter used to pick off the bright-port signal, which is collected by PD1. Two additional photodetectors, PD2 and PD3, monitor the DC power at the dark-port and the cavity transmission, respectively.

The common mode $L_{+}$ of the L-shaped cavity is controlled by synchronously actuating the two ETMs using piezoelectric actuators. Once $L_{+}$ is locked to the laser frequency, the differential mode $\mathcal{L}_-$ can be scanned either by modulating the steering mirror in the Sagnac loop $l_-$ or by differentially actuating the two ETMs $L_-$, consistent with the degeneracy in Eq.\,\eqref{eq:darkDC and brightDC when re=1}. In the experiment, we adopt the former approach by applying a ramp signal to the piezoelectric actuator of the Sagnac steering mirror. During the scan, the corresponding DC powers at all three ports were recorded in real time.

\begin{figure}[h!]
    \centering
    
  \subfloat[]{
        \centering    \includegraphics[width=0.9\linewidth]{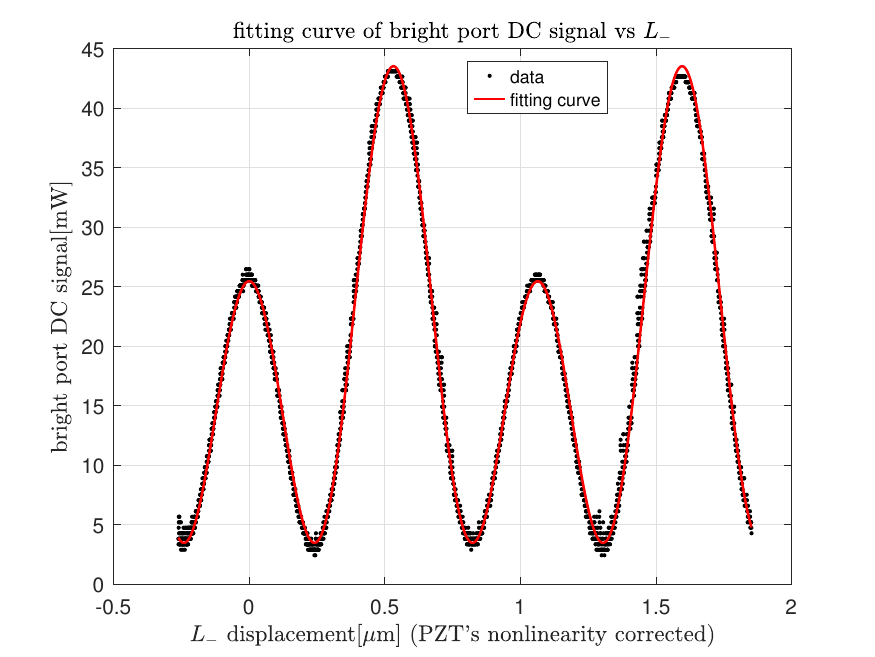}      \label{fig:bright}
        }

    \vspace{0.1em} 
    
  \subfloat[]{
        \centering \includegraphics[width=0.9\linewidth]{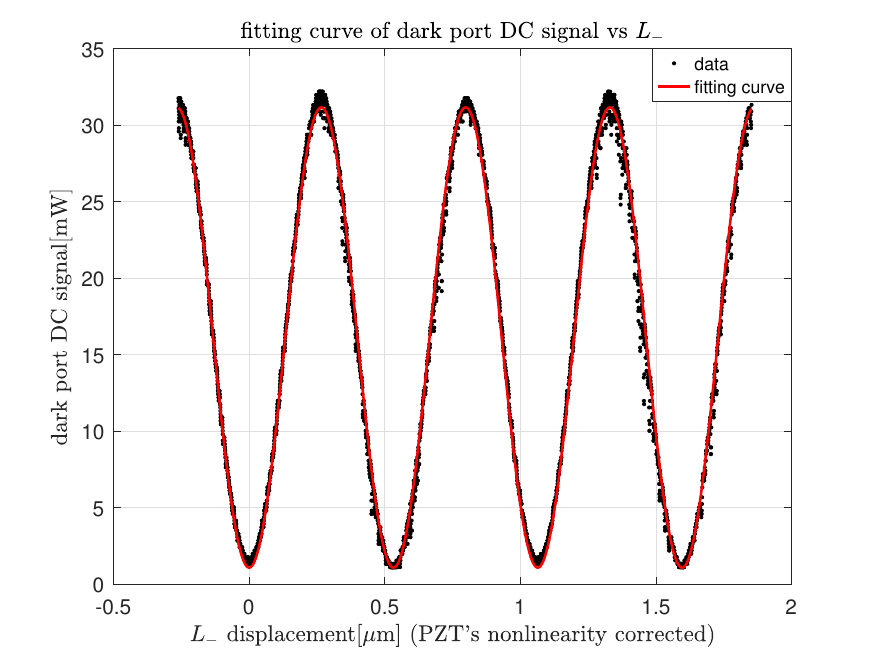}     \label{fig:dark}
        }
    
    \vspace{0.1em}
    
  \subfloat[]{
        \centering \includegraphics[width=0.9\linewidth]{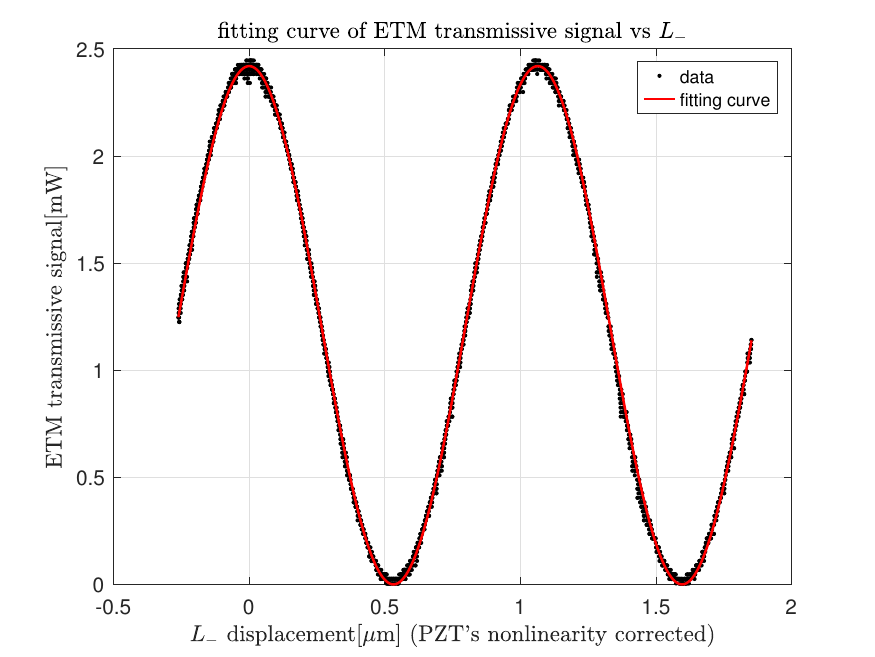}    \label{fig:trans}
         }
    \caption{Measured output power at the (a) bright, (b) dark, and (c) transmission ports. Data were fitted based on Eqs.\,\eqref{eq:darkDC when re<1}-\eqref{eq:intra-cavityDC when re<1}, yielding reflectivities of $R_{\mathrm{i}} = 97.8\%$ for the ITM and $R_{\mathrm{e}} = 99.7\%$ for two ETMs.}
    \label{fig:result}
\end{figure}

For quantitative comparison with the theoretical model, the following calibration procedures were performed.
Firstly, the photodetector signals were converted to optical power, taking into account variations in detector responsivity and the optical transmission factors. Secondly, the piezoelectric actuator response was independently calibrated, and its intrinsic nonlinearity was corrected. This procedure linearized the scan, enabling a direct conversion of the actuator drive voltage to an equivalent physical displacement. Finally, the calibrated signals from the photodetectors were fitted simultaneously to the analytic expressions in Eqs.\,\eqref{eq:darkDC when re<1}\,-\,\eqref{eq:intra-cavityDC when re<1}, which describe the dependence of the bright-port, dark-port, and transmission port powers on the differential length.
Figure\,\ref{fig:result} shows the calibrated data overlaid with the best-fit theoretical curves. We have used two fitting parameters: the reflectivities of the ITM and ETM. The best-fit values,  $R_{\rm ITM} = 97.8\%$ and $R_{\rm ETM} = 99.7\%$, agree well with the specifications from the optics vendor. 

There are overall offsets observed in the bright-port and dark-port signals, which were not covered by the idealized theoretical model. The model predicts that all three DC signals should reach zero at specific values of the differential length. This deviation arises because the analytical derivation assumes ideal plane-wave propagation and neglects higher-order transverse modes. In practice, residual mode mismatch and imperfect alignment introduce contrast defects at the output ports that lead to nonzero offsets. To assess the impact of mode mismatch more quantitatively, we have carried out a numerical  simulation using Finesse\,\cite{brown_2025_12662017}. We found a mode mismatch of order of 4\% is sufficient to reproduce the observed offsets, which agrees with the difference between the measured beam waist near ITM and the value expected from the cavity geometry. The full set of experimental parameters is summarized in Table\,\ref{tb:para1}.

\begin{table}[t]
\centering
\begin{tabular}{|l|l|}
\hline
\textbf{Parameter} & \textbf{Value}\\
\hline
EOM modulation frequency & 15\,MHz\\
EOM modulation depth & 0.15 rad\\
cavity arm common mode $L_+$ & 20.3\,cm\\
cavity arm differential mode $L_-$ &  0.1 \,cm \\
Sagnac common mode $l_+$& 46.3\,cm \\
Sagnac differential mode $l_-$& 0.3 \,cm \\
ITM reflection $R_{\rm i}$ & 97.8\%\\
ETM reflection $R_{\rm e}$ &  99.7\%\\
ETM radius of curvature & 1\,m \\
ideal beam waist radius at ITM& 318.8\,$\mu$m\\
measured beam waist radius at ITM& 304.5\,$\mu$m\\
beam waist radius mismatch & 14.3\,$\mu$m\\
beam waist position mismatch & 11.1\,cm\\
mode mismatch at ITM & 3.85\%\\
\hline
\end{tabular}
\caption{Parameters of the experiment. The reflection of the mode mismatch is given directly by Finesse.}
\label{tb:para1}
\end{table}

\section{conclusion}\label{sec:summary}
In this work, we have investigated the DC optical response of the recently proposed L-shaped cavity interferometer topology for kilohertz gravitational-wave detection. Through a controlled tabletop experiment, we verified two key theoretical predictions: (i) when the laser is locked to the resonance of the L-shaped cavity, the prompt and first-pass reflection cancel in such a way that the cavity input coupler becomes effectively transparent, and the interferometer with an L-shaped cavity behaves as a folded Michelson; and (ii) the Sagnac vortex decomposes into two independent pumping paths whose interference governs both the intra-cavity power and the output-port signals. The experimental results, obtained through calibrated scans of the differential degree of freedom, show nice agreement with the analytic and numerical model. The intuitive physical picture developed here—namely, the transparent-ITM behavior at resonance, the degeneracy between the cavity and Sagnac differential modes, and the two-path pumping interpretation—helps clarify the lock-acquisition pathways and the structure of error signals available for stabilizing the interferometer in its target operating state. 

Looking forward, these results form a stepping stone toward implementing this topology in larger-scale prototypes, e.g., the one under construction at Beijing Normal University\,\cite{wang26}, realizing a suspended interferometer with the L-shaped cavity. In the longer term, the insights developed here support the broader program of engineering next-generation kHz gravitational-wave detectors capable of probing neutron-star post-merger oscillations. As the L-shaped cavity topology is further integrated with squeezed-light injection, high-power operation, and low-loss optical coatings, the combination of theoretical modeling, tabletop validation, and intermediate-scale prototypes will help assess its viability for future observatories.

\section{Acknowledge}
This work was supported by the National Key Research and Development Program of China (Grant Nos. 2024YFC2208000, and 2023YFC2205800), the National Natural Science Foundation of China (Grant Nos. 12441503 and 12433001), the Fundamental Research Funds for the Central Universities of China(Grant No. 310432103). We also acknowledge the support of the Institute for Gravitational Wave Astronomy at the University of Birmingham, UK.

\bibliography{ref}

@article{wang26,
	author = {Wang, Mengyao and Zhang, Fan and Guo, Xinyao and Miao, Haixing and Yang, Huan and Ma, Yiqiu and Wang, Haoyu and Wang, Haibo and Zhang, Teng and Cao, Mengdi and Chen, Yuchao and Huang, Xiaoman and Liu, Fangfei and Liu, Jianyu and Pan, Yuan and Li, Junlang and Xia, Yulin and Xing, Jianbo and Yu, Yujie and Zhou, Chenjie and Zhu, Zonghong},
	date = {2026/01/04},
	doi = {10.1007/s11433-025-2831-1},
	id = {Wang2026},
	isbn = {1869-1927},
	journal = {Science China Physics, Mechanics \& Astronomy},
	number = {3},
	pages = {239511},
	title = {Beijing Normal University 12-meter interferometric kHz gravitational wave detector prototype: Design and scientific prospects},
	url = {https://doi.org/10.1007/s11433-025-2831-1},
	volume = {69},
	year = {2026}}

@article{Vinet,
  title = {Optimization of long-baseline optical interferometers for gravitational-wave detection},
  author = {Vinet, Jean-Yves and Meers, Brian and Man, Catherine Nary and Brillet, Alain},
  journal = {Phys. Rev. D},
  volume = {38},
  issue = {2},
  pages = {433--447},
  numpages = {0},
  year = {1988},
  month = {Jul},
  publisher = {American Physical Society},
  doi = {10.1103/PhysRevD.38.433},
  url = {https://link.aps.org/doi/10.1103/PhysRevD.38.433}
}

@article{Synchronous,
    author = {R. W. Drever},
    title = {Interferometric Detectors for Gravitational
Radiation},
    journal = {Lect. Notes Phys. 124, 321},
    year = {1983}
}

@article{Smith,
    author = {P. Smith},
    title = {Stabilized, single-frequency output from a
long laser cavity},
    journal = {IEEE Journal of Quantum Electronics,
1(8):343–348},
    year = {1965}
}

@phdthesis{Evansthesis,
    author       = {Matthew Evans},
    title        = {Lock Acquisition in Resonant
Optical Interferometers},
    school       = {California Institute of Technology},
    year         = {2002}
}

@phdthesis{Martynovthesis,
    author       = {Denis Martynov},
    title        = {Lock Acquisition and Sensitivity Analysis of Advanced LIGO Interferometers},
    school       = {California Institute of Technology},
    year         = {2015}
}

@article{Staley2014,
doi = {10.1088/0264-9381/31/24/245010},
url = {https://doi.org/10.1088/0264-9381/31/24/245010},
year = {2014},
month = {nov},
publisher = {IOP Publishing},
volume = {31},
number = {24},
pages = {245010},
author = {Staley, A and Martynov, D and Abbott, R and Adhikari, R X and Arai, K and Ballmer, S and Barsotti, L and Brooks, A F and DeRosa, R T and Dwyer, S and Effler, A and Evans, M and Fritschel, P and Frolov, V V and Gray, C and Guido, C J and Gustafson, R and Heintze, M and Hoak, D and Izumi, K and Kawabe, K and King, E J and Kissel, J S and Kokeyama, K and Landry, M and McClelland, D E and Miller, J and Mullavey, A and OʼReilly, B and Rollins, J G and Sanders, J R and Schofield, R M S and Sigg, D and Slagmolen, B J J and Smith-Lefebvre, N D and Vajente, G and Ward, R L and Wipf, C},
title = {Achieving resonance in the Advanced LIGO gravitational-wave interferometer},
journal = {Classical and Quantum Gravity}
}

@article{Miao19,
  title = {Quantum Limit for Laser Interferometric Gravitational-Wave Detectors from Optical Dissipation},
  author = {Miao, Haixing and Smith, Nicolas D. and Evans, Matthew},
  journal = {Phys. Rev. X},
  volume = {9},
  issue = {1},
  pages = {011053},
  numpages = {10},
  year = {2019},
  month = {Mar},
  publisher = {American Physical Society},
  doi = {10.1103/PhysRevX.9.011053},
  url = {https://link.aps.org/doi/10.1103/PhysRevX.9.011053}
}

@article{O5upgrade,
    author = {LSC memeber},
    title = {Report of the LSC Post-O5} ,
    journal = {LIGO Technical notes-T2200287},
    year = {2023},
url = {https://dcc-lho.ligo.org/LIGO-T2200287/public}

}

@article{NEMO2020,
title={Neutron Star Extreme Matter Observatory: A kilohertz-band gravitational-wave detector in the global network}, volume={37}, DOI={10.1017/pasa.2020.39}, journal={Publications of the Astronomical Society of Australia}, author={Ackley, K. and Adya, V. B. and Agrawal, P. and Altin, P. and Ashton, G. and Bailes, M. and Baltinas, E. and Barbuio, A. and Beniwal, D. and Blair, C. and et al.}, year={2020}, pages={e047}}

@article{Miao18,
  title = {Towards the design of gravitational-wave detectors for probing neutron-star physics},
  author = {Miao, Haixing and Yang, Huan and Martynov, Denis},
  journal = {Phys. Rev. D},
  volume = {98},
  issue = {4},
  pages = {044044},
  numpages = {15},
  year = {2018},
  month = {Aug},
  publisher = {American Physical Society},
  doi = {10.1103/PhysRevD.98.044044},
  url = {https://link.aps.org/doi/10.1103/PhysRevD.98.044044}
}

@article{Martynov19,
  title = {Exploring the sensitivity of gravitational wave detectors to neutron star physics},
  author = {Martynov, Denis and Miao, Haixing and Yang, Huan and Vivanco, Francisco Hernandez and Thrane, Eric and Smith, Rory and Lasky, Paul and East, William E. and Adhikari, Rana and Bauswein, Andreas and Brooks, Aidan and Chen, Yanbei and Corbitt, Thomas and Freise, Andreas and Grote, Hartmut and Levin, Yuri and Zhao, Chunnong and Vecchio, Alberto},
  journal = {Phys. Rev. D},
  volume = {99},
  issue = {10},
  pages = {102004},
  numpages = {17},
  year = {2019},
  month = {May},
  publisher = {American Physical Society},
  doi = {10.1103/PhysRevD.99.102004},
  url = {https://link.aps.org/doi/10.1103/PhysRevD.99.102004}
}

@article{DanilishinLivingReview,
	author = {Danilishin, Stefan L. and Khalili, Farid Ya. and Miao, Haixing},
	date = {2019/04/29},
	doi = {10.1007/s41114-019-0018-y},
	id = {Danilishin2019},
	isbn = {1433-8351},
	journal = {Living Reviews in Relativity},
	number = {1},
	pages = {2},
	title = {Advanced quantum techniques for future gravitational-wave detectors},
	url = {https://doi.org/10.1007/s41114-019-0018-y},
	volume = {22},
	year = {2019}}

@article{RanaRMP,
  title = {Gravitational radiation detection with laser interferometry},
  author = {Adhikari, Rana X.},
  journal = {Rev. Mod. Phys.},
  volume = {86},
  issue = {1},
  pages = {121--151},
  numpages = {0},
  year = {2014},
  month = {Feb},
  publisher = {American Physical Society},
  doi = {10.1103/RevModPhys.86.121},
  url = {https://link.aps.org/doi/10.1103/RevModPhys.86.121}
}

@article{Chen_2013,
doi = {10.1088/0953-4075/46/10/104001},
url = {https://doi.org/10.1088/0953-4075/46/10/104001},
year = {2013},
month = {may},
publisher = {IOP Publishing},
volume = {46},
number = {10},
pages = {104001},
author = {Chen, Yanbei},
title = {Macroscopic quantum mechanics: theory and experimental concepts of optomechanics},
journal = {Journal of Physics B: Atomic, Molecular and Optical Physics}
}

@ARTICLE{GWTC-3,
       author = {{Abbott}, R. and {Abbott}, T.~D. and {Acernese}, F. and {Ackley}, K. and {Adams}, C. and {Adhikari}, N. and {Adhikari}, R.~X. and {Adya}, V.~B. and {Affeldt}, C. and {Agarwal}, D. and {Agathos}, M. and {Agatsuma}, K. and {Aggarwal}, N. and {Aguiar}, O.~D. and {Aiello}, L. and {Ain}, A. and {Ajith}, P. and {Akcay}, S. and {Akutsu}, T. and {Albanesi}, S. and {Allocca}, A. and {Altin}, P.~A. and {Amato}, A. and {Anand}, C. and {Anand}, S. and {Ananyeva}, A. and {Anderson}, S.~B. and {Anderson}, W.~G. and {Ando}, M. and {Andrade}, T. and {Andres}, N. and {Andri{\'c}}, T. and {Angelova}, S.~V. and {Ansoldi}, S. and {Antelis}, J.~M. and {Antier}, S. and {Appert}, S. and {Arai}, Koji and {Arai}, Koya and {Arai}, Y. and {Araki}, S. and {Araya}, A. and {Araya}, M.~C. and {Areeda}, J.~S. and {Ar{\`e}ne}, M. and {Aritomi}, N. and {Arnaud}, N. and {Arogeti}, M. and {Aronson}, S.~M. and {Arun}, K.~G. and {Asada}, H. and {Asali}, Y. and {Ashton}, G. and {Aso}, Y. and {Assiduo}, M. and {Aston}, S.~M. and {Astone}, P. and {Aubin}, F. and {Austin}, C. and {Babak}, S. and {Badaracco}, F. and {Bader}, M.~K.~M. and {Badger}, C. and {Bae}, S. and {Bae}, Y. and {Baer}, A.~M. and {Bagnasco}, S. and {Bai}, Y. and {Baiotti}, L. and {Baird}, J. and {Bajpai}, R. and {Ball}, M. and {Ballardin}, G. and {Ballmer}, S.~W. and {Balsamo}, A. and {Baltus}, G. and {Banagiri}, S. and {Bankar}, D. and {Barayoga}, J.~C. and {Barbieri}, C. and {Barish}, B.~C. and {Barker}, D. and {Barneo}, P. and {Barone}, F. and {Barr}, B. and {Barsotti}, L. and {Barsuglia}, M. and {Barta}, D. and {Bartlett}, J. and {Barton}, M.~A. and {Bartos}, I. and {Bassiri}, R. and {Basti}, A. and {Bawaj}, M. and {Bayley}, J.~C. and {Baylor}, A.~C. and {Bazzan}, M. and {B{\'e}csy}, B. and {Bedakihale}, V.~M. and {Bejger}, M. and {Belahcene}, I. and {Benedetto}, V. and {Beniwal}, D. and {Bennett}, T.~F. and {Bentley}, J.~D. and {Benyaala}, M. and {Bergamin}, F. and {Berger}, B.~K. and {Bernuzzi}, S. and {Berry}, C.~P.~L. and {Bersanetti}, D. and {Bertolini}, A. and {Betzwieser}, J. and {Beveridge}, D. and {Bhandare}, R. and {Bhardwaj}, U. and {Bhattacharjee}, D. and {Bhaumik}, S. and {Bilenko}, I.~A. and {Billingsley}, G. and {Bini}, S. and {Birney}, R. and {Birnholtz}, O. and {Biscans}, S. and {Bischi}, M. and {Biscoveanu}, S. and {Bisht}, A. and {Biswas}, B. and {Bitossi}, M. and {Bizouard}, M.-A. and {Blackburn}, J.~K. and {Blair}, C.~D. and {Blair}, D.~G. and {Blair}, R.~M. and {Bobba}, F. and {Bode}, N. and {Boer}, M. and {Bogaert}, G. and {Boldrini}, M. and {Bonavena}, L.~D. and {Bondu}, F. and {Bonilla}, E. and {Bonnand}, R. and {Booker}, P. and {Boom}, B.~A. and {Bork}, R. and {Boschi}, V. and {Bose}, N. and {Bose}, S. and {Bossilkov}, V. and {Boudart}, V. and {Bouffanais}, Y. and {Bozzi}, A. and {Bradaschia}, C. and {Brady}, P.~R. and {Bramley}, A. and {Branch}, A. and {Branchesi}, M. and {Brandt}, J. and {Brau}, J.~E. and {Breschi}, M. and {Briant}, T. and {Briggs}, J.~H. and {Brillet}, A. and {Brinkmann}, M. and {Brockill}, P. and {Brooks}, A.~F. and {Brooks}, J. and {Brown}, D.~D. and {Brunett}, S. and {Bruno}, G. and {Bruntz}, R. and {Bryant}, J. and {Bulik}, T. and {Bulten}, H.~J. and {Buonanno}, A. and {Buscicchio}, R. and {Buskulic}, D. and {Buy}, C. and {Byer}, R.~L. and {Davies}, G.~S. Cabourn and {Cadonati}, L. and {Cagnoli}, G. and {Cahillane}, C. and {Bustillo}, J. Calder{\'o}n and {Callaghan}, J.~D. and {Callister}, T.~A. and {Calloni}, E. and {Cameron}, J. and {Camp}, J.~B. and {Canepa}, M. and {Canevarolo}, S. and {Cannavacciuolo}, M. and {Cannon}, K.~C. and {Cao}, H. and {Cao}, Z. and {Capocasa}, E. and {Capote}, E. and {Carapella}, G. and {Carbognani}, F.},
        title = "{GWTC-3: Compact Binary Coalescences Observed by LIGO and Virgo during the Second Part of the Third Observing Run}",
      journal = {Physical Review X},
         year = 2023,
        month = oct,
       volume = {13},
       number = {4},
          eid = {041039},
        pages = {041039},
          doi = {10.1103/PhysRevX.13.041039},
archivePrefix = {arXiv},
       eprint = {2111.03606},
 primaryClass = {gr-qc},
       adsurl = {https://ui.adsabs.harvard.edu/abs/2023PhRvX..13d1039A}
}

@misc{brown_2025_12662017,
  author       = {Brown, Daniel David and
                  Freise, Andreas and
                  Cao, Huy Tuong and
                  Ciobanu, Alexei and
                  Gobeil, Jeremie and
                  Green, Anna and
                  Hapke, Paul and
                  Jones, Philip and
                  van der Kolk, Miron and
                  Kuns, Kevin and
                  Leavey, Sean and
                  Perry, Jonathan Warren and
                  Rowlinson, Samuel and
                  Sallé, Mischa},
  title        = {FINESSE},
  month        = mar,
  year         = 2025,
  publisher    = {Gitlab},
  version      = {3.0a32},
  doi          = {10.5281/zenodo.12662017},
  url          = {https://doi.org/10.5281/zenodo.12662017},
}

@article{PhysRevLett.116.061102,
  title = {Observation of Gravitational Waves from a Binary Black Hole Merger},
  author = {Abbott, B. P. and Abbott, R. and Abbott, T. D. and Abernathy, M. R. and Acernese, F. and Ackley, K. and Adams, C. and Adams, T. and Addesso, P. and Adhikari, R. X. and Adya, V. B. and Affeldt, C. and Agathos, M. and Agatsuma, K. and Aggarwal, N. and Aguiar, O. D. and Aiello, L. and Ain, A. and Ajith, P. and Allen, B. and Allocca, A. and Altin, P. A. and Anderson, S. B. and Anderson, W. G. and Arai, K. and Arain, M. A. and Araya, M. C. and Arceneaux, C. C. and Areeda, J. S. and Arnaud, N. and Arun, K. G. and Ascenzi, S. and Ashton, G. and Ast, M. and Aston, S. M. and Astone, P. and Aufmuth, P. and Aulbert, C. and Babak, S. and Bacon, P. and Bader, M. K. M. and Baker, P. T. and Baldaccini, F. and Ballardin, G. and Ballmer, S. W. and Barayoga, J. C. and Barclay, S. E. and Barish, B. C. and Barker, D. and Barone, F. and Barr, B. and Barsotti, L. and Barsuglia, M. and Barta, D. and Bartlett, J. and Barton, M. A. and Bartos, I. and Bassiri, R. and Basti, A. and Batch, J. C. and Baune, C. and Bavigadda, V. and Bazzan, M. and Behnke, B. and Bejger, M. and Belczynski, C. and Bell, A. S. and Bell, C. J. and Berger, B. K. and Bergman, J. and Bergmann, G. and Berry, C. P. L. and Bersanetti, D. and Bertolini, A. and Betzwieser, J. and Bhagwat, S. and Bhandare, R. and Bilenko, I. A. and Billingsley, G. and Birch, J. and Birney, I. A. and Birnholtz, O. and Biscans, S. and Bisht, A. and Bitossi, M. and Biwer, C. and Bizouard, M. A. and Blackburn, J. K. and Blair, C. D. and Blair, D. G. and Blair, R. M. and Bloemen, S. and Bock, O. and Bodiya, T. P. and Boer, M. and Bogaert, G. and Bogan, C. and Bohe, A. and Bojtos, P. and Bond, C. and Bondu, F. and Bonnand, R. and Boom, B. A. and Bork, R. and Boschi, V. and Bose, S. and Bouffanais, Y. and Bozzi, A. and Bradaschia, C. and Brady, P. R. and Braginsky, V. B. and Branchesi, M. and Brau, J. E. and Briant, T. and Brillet, A. and Brinkmann, M. and Brisson, V. and Brockill, P. and Brooks, A. F. and Brown, D. A. and Brown, D. D. and Brown, N. M. and Buchanan, C. C. and Buikema, A. and Bulik, T. and Bulten, H. J. and Buonanno, A. and Buskulic, D. and Buy, C. and Byer, R. L. and Cabero, M. and Cadonati, L. and Cagnoli, G. and Cahillane, C. and Bustillo, J. Calder\'on and Callister, T. and Calloni, E. and Camp, J. B. and Cannon, K. C. and Cao, J. and Capano, C. D. and Capocasa, E. and Carbognani, F. and Caride, S. and Diaz, J. Casanueva and Casentini, C. and Caudill, S. and Cavagli\`a, M. and Cavalier, F. and Cavalieri, R. and Cella, G. and Cepeda, C. B. and Baiardi, L. Cerboni and Cerretani, G. and Cesarini, E. and Chakraborty, R. and Chalermsongsak, T. and Chamberlin, S. J. and Chan, M. and Chao, S. and Charlton, P. and Chassande-Mottin, E. and Chen, H. Y. and Chen, Y. and Cheng, C. and Chincarini, A. and Chiummo, A. and Cho, H. S. and Cho, M. and Chow, J. H. and Christensen, N. and Chu, Q. and Chua, S. and Chung, S. and Ciani, G. and Clara, F. and Clark, J. A. and Cleva, F. and Coccia, E. and Cohadon, P.-F. and Colla, A. and Collette, C. G. and Cominsky, L. and Constancio, M. and Conte, A. and Conti, L. and Cook, D. and Corbitt, T. R. and Cornish, N. and Corsi, A. and Cortese, S. and Costa, C. A. and Coughlin, M. W. and Coughlin, S. B. and Coulon, J.-P. and Countryman, S. T. and Couvares, P. and Cowan, E. E. and Coward, D. M. and Cowart, M. J. and Coyne, D. C. and Coyne, R. and Craig, K. and Creighton, J. D. E. and Creighton, T. D. and Cripe, J. and Crowder, S. G. and Cruise, A. M. and Cumming, A. and Cunningham, L. and Cuoco, E. and Canton, T. Dal and Danilishin, S. L. and D'Antonio, S. and Danzmann, K. and Darman, N. S. and Da Silva Costa, C. F. and Dattilo, V. and Dave, I. and Daveloza, H. P. and Davier, M. and Davies, G. S. and Daw, E. J. and Day, R. and De, S. and DeBra, D. and Debreczeni, G. and Degallaix, J. and De Laurentis, M. and Del\'eglise, S. and Del Pozzo, W. and Denker, T. and Dent, T. and Dereli, H. and Dergachev, V. and DeRosa, R. T. and De Rosa, R. and DeSalvo, R. and Dhurandhar, S. and D\'{\i}az, M. C. and Di Fiore, L. and Di Giovanni, M. and Di Lieto, A. and Di Pace, S. and Di Palma, I. and Di Virgilio, A. and Dojcinoski, G. and Dolique, V. and Donovan, F. and Dooley, K. L. and Doravari, S. and Douglas, R. and Downes, T. P. and Drago, M. and Drever, R. W. P. and Driggers, J. C. and Du, Z. and Ducrot, M. and Dwyer, S. E. and Edo, T. B. and Edwards, M. C. and Effler, A. and Eggenstein, H.-B. and Ehrens, P. and Eichholz, J. and Eikenberry, S. S. and Engels, W. and Essick, R. C. and Etzel, T. and Evans, M. and Evans, T. M. and Everett, R. and Factourovich, M. and Fafone, V. and Fair, H. and Fairhurst, S. and Fan, X. and Fang, Q. and Farinon, S. and Farr, B. and Farr, W. M. and Favata, M. and Fays, M. and Fehrmann, H. and Fejer, M. M. and Feldbaum, D. and Ferrante, I. and Ferreira, E. C. and Ferrini, F. and Fidecaro, F. and Finn, L. S. and Fiori, I. and Fiorucci, D. and Fisher, R. P. and Flaminio, R. and Fletcher, M. and Fong, H. and Fournier, J.-D. and Franco, S. and Frasca, S. and Frasconi, F. and Frede, M. and Frei, Z. and Freise, A. and Frey, R. and Frey, V. and Fricke, T. T. and Fritschel, P. and Frolov, V. V. and Fulda, P. and Fyffe, M. and Gabbard, H. A. G. and Gair, J. R. and Gammaitoni, L. and Gaonkar, S. G. and Garufi, F. and Gatto, A. and Gaur, G. and Gehrels, N. and Gemme, G. and Gendre, B. and Genin, E. and Gennai, A. and George, J. and Gergely, L. and Germain, V. and Ghosh, Abhirup and Ghosh, Archisman and Ghosh, S. and Giaime, J. A. and Giardina, K. D. and Giazotto, A. and Gill, K. and Glaefke, A. and Gleason, J. R. and Goetz, E. and Goetz, R. and Gondan, L. and Gonz\'alez, G. and Castro, J. M. Gonzalez and Gopakumar, A. and Gordon, N. A. and Gorodetsky, M. L. and Gossan, S. E. and Gosselin, M. and Gouaty, R. and Graef, C. and Graff, P. B. and Granata, M. and Grant, A. and Gras, S. and Gray, C. and Greco, G. and Green, A. C. and Greenhalgh, R. J. S. and Groot, P. and Grote, H. and Grunewald, S. and Guidi, G. M. and Guo, X. and Gupta, A. and Gupta, M. K. and Gushwa, K. E. and Gustafson, E. K. and Gustafson, R. and Hacker, J. J. and Hall, B. R. and Hall, E. D. and Hammond, G. and Haney, M. and Hanke, M. M. and Hanks, J. and Hanna, C. and Hannam, M. D. and Hanson, J. and Hardwick, T. and Harms, J. and Harry, G. M. and Harry, I. W. and Hart, M. J. and Hartman, M. T. and Haster, C.-J. and Haughian, K. and Healy, J. and Heefner, J. and Heidmann, A. and Heintze, M. C. and Heinzel, G. and Heitmann, H. and Hello, P. and Hemming, G. and Hendry, M. and Heng, I. S. and Hennig, J. and Heptonstall, A. W. and Heurs, M. and Hild, S. and Hoak, D. and Hodge, K. A. and Hofman, D. and Hollitt, S. E. and Holt, K. and Holz, D. E. and Hopkins, P. and Hosken, D. J. and Hough, J. and Houston, E. A. and Howell, E. J. and Hu, Y. M. and Huang, S. and Huerta, E. A. and Huet, D. and Hughey, B. and Husa, S. and Huttner, S. H. and Huynh-Dinh, T. and Idrisy, A. and Indik, N. and Ingram, D. R. and Inta, R. and Isa, H. N. and Isac, J.-M. and Isi, M. and Islas, G. and Isogai, T. and Iyer, B. R. and Izumi, K. and Jacobson, M. B. and Jacqmin, T. and Jang, H. and Jani, K. and Jaranowski, P. and Jawahar, S. and Jim\'enez-Forteza, F. and Johnson, W. W. and Johnson-McDaniel, N. K. and Jones, D. I. and Jones, R. and Jonker, R. J. G. and Ju, L. and Haris, K. and Kalaghatgi, C. V. and Kalogera, V. and Kandhasamy, S. and Kang, G. and Kanner, J. B. and Karki, S. and Kasprzack, M. and Katsavounidis, E. and Katzman, W. and Kaufer, S. and Kaur, T. and Kawabe, K. and Kawazoe, F. and K\'ef\'elian, F. and Kehl, M. S. and Keitel, D. and Kelley, D. B. and Kells, W. and Kennedy, R. and Keppel, D. G. and Key, J. S. and Khalaidovski, A. and Khalili, F. Y. and Khan, I. and Khan, S. and Khan, Z. and Khazanov, E. A. and Kijbunchoo, N. and Kim, C. and Kim, J. and Kim, K. and Kim, Nam-Gyu and Kim, Namjun and Kim, Y.-M. and King, E. J. and King, P. J. and Kinzel, D. L. and Kissel, J. S. and Kleybolte, L. and Klimenko, S. and Koehlenbeck, S. M. and Kokeyama, K. and Koley, S. and Kondrashov, V. and Kontos, A. and Koranda, S. and Korobko, M. and Korth, W. Z. and Kowalska, I. and Kozak, D. B. and Kringel, V. and Krishnan, B. and Kr\'olak, A. and Krueger, C. and Kuehn, G. and Kumar, P. and Kumar, R. and Kuo, L. and Kutynia, A. and Kwee, P. and Lackey, B. D. and Landry, M. and Lange, J. and Lantz, B. and Lasky, P. D. and Lazzarini, A. and Lazzaro, C. and Leaci, P. and Leavey, S. and Lebigot, E. O. and Lee, C. H. and Lee, H. K. and Lee, H. M. and Lee, K. and Lenon, A. and Leonardi, M. and Leong, J. R. and Leroy, N. and Letendre, N. and Levin, Y. and Levine, B. M. and Li, T. G. F. and Libson, A. and Littenberg, T. B. and Lockerbie, N. A. and Logue, J. and Lombardi, A. L. and London, L. T. and Lord, J. E. and Lorenzini, M. and Loriette, V. and Lormand, M. and Losurdo, G. and Lough, J. D. and Lousto, C. O. and Lovelace, G. and L\"uck, H. and Lundgren, A. P. and Luo, J. and Lynch, R. and Ma, Y. and MacDonald, T. and Machenschalk, B. and MacInnis, M. and Macleod, D. M. and Maga\~na-Sandoval, F. and Magee, R. M. and Mageswaran, M. and Majorana, E. and Maksimovic, I. and Malvezzi, V. and Man, N. and Mandel, I. and Mandic, V. and Mangano, V. and Mansell, G. L. and Manske, M. and Mantovani, M. and Marchesoni, F. and Marion, F. and M\'arka, S. and M\'arka, Z. and Markosyan, A. S. and Maros, E. and Martelli, F. and Martellini, L. and Martin, I. W. and Martin, R. M. and Martynov, D. V. and Marx, J. N. and Mason, K. and Masserot, A. and Massinger, T. J. and Masso-Reid, M. and Matichard, F. and Matone, L. and Mavalvala, N. and Mazumder, N. and Mazzolo, G. and McCarthy, R. and McClelland, D. E. and McCormick, S. and McGuire, S. C. and McIntyre, G. and McIver, J. and McManus, D. J. and McWilliams, S. T. and Meacher, D. and Meadors, G. D. and Meidam, J. and Melatos, A. and Mendell, G. and Mendoza-Gandara, D. and Mercer, R. A. and Merilh, E. and Merzougui, M. and Meshkov, S. and Messenger, C. and Messick, C. and Meyers, P. M. and Mezzani, F. and Miao, H. and Michel, C. and Middleton, H. and Mikhailov, E. E. and Milano, L. and Miller, J. and Millhouse, M. and Minenkov, Y. and Ming, J. and Mirshekari, S. and Mishra, C. and Mitra, S. and Mitrofanov, V. P. and Mitselmakher, G. and Mittleman, R. and Moggi, A. and Mohan, M. and Mohapatra, S. R. P. and Montani, M. and Moore, B. C. and Moore, C. J. and Moraru, D. and Moreno, G. and Morriss, S. R. and Mossavi, K. and Mours, B. and Mow-Lowry, C. M. and Mueller, C. L. and Mueller, G. and Muir, A. W. and Mukherjee, Arunava and Mukherjee, D. and Mukherjee, S. and Mukund, N. and Mullavey, A. and Munch, J. and Murphy, D. J. and Murray, P. G. and Mytidis, A. and Nardecchia, I. and Naticchioni, L. and Nayak, R. K. and Necula, V. and Nedkova, K. and Nelemans, G. and Neri, M. and Neunzert, A. and Newton, G. and Nguyen, T. T. and Nielsen, A. B. and Nissanke, S. and Nitz, A. and Nocera, F. and Nolting, D. and Normandin, M. E. N. and Nuttall, L. K. and Oberling, J. and Ochsner, E. and O'Dell, J. and Oelker, E. and Ogin, G. H. and Oh, J. J. and Oh, S. H. and Ohme, F. and Oliver, M. and Oppermann, P. and Oram, Richard J. and O'Reilly, B. and O'Shaughnessy, R. and Ott, C. D. and Ottaway, D. J. and Ottens, R. S. and Overmier, H. and Owen, B. J. and Pai, A. and Pai, S. A. and Palamos, J. R. and Palashov, O. and Palomba, C. and Pal-Singh, A. and Pan, H. and Pan, Y. and Pankow, C. and Pannarale, F. and Pant, B. C. and Paoletti, F. and Paoli, A. and Papa, M. A. and Paris, H. R. and Parker, W. and Pascucci, D. and Pasqualetti, A. and Passaquieti, R. and Passuello, D. and Patricelli, B. and Patrick, Z. and Pearlstone, B. L. and Pedraza, M. and Pedurand, R. and Pekowsky, L. and Pele, A. and Penn, S. and Perreca, A. and Pfeiffer, H. P. and Phelps, M. and Piccinni, O. and Pichot, M. and Pickenpack, M. and Piergiovanni, F. and Pierro, V. and Pillant, G. and Pinard, L. and Pinto, I. M. and Pitkin, M. and Poeld, J. H. and Poggiani, R. and Popolizio, P. and Post, A. and Powell, J. and Prasad, J. and Predoi, V. and Premachandra, S. S. and Prestegard, T. and Price, L. R. and Prijatelj, M. and Principe, M. and Privitera, S. and Prix, R. and Prodi, G. A. and Prokhorov, L. and Puncken, O. and Punturo, M. and Puppo, P. and P\"urrer, M. and Qi, H. and Qin, J. and Quetschke, V. and Quintero, E. A. and Quitzow-James, R. and Raab, F. J. and Rabeling, D. S. and Radkins, H. and Raffai, P. and Raja, S. and Rakhmanov, M. and Ramet, C. R. and Rapagnani, P. and Raymond, V. and Razzano, M. and Re, V. and Read, J. and Reed, C. M. and Regimbau, T. and Rei, L. and Reid, S. and Reitze, D. H. and Rew, H. and Reyes, S. D. and Ricci, F. and Riles, K. and Robertson, N. A. and Robie, R. and Robinet, F. and Rocchi, A. and Rolland, L. and Rollins, J. G. and Roma, V. J. and Romano, J. D. and Romano, R. and Romanov, G. and Romie, J. H. and Rosi\ifmmode \acute{n}\else \'{n}\fi{}ska, D. and Rowan, S. and R\"udiger, A. and Ruggi, P. and Ryan, K. and Sachdev, S. and Sadecki, T. and Sadeghian, L. and Salconi, L. and Saleem, M. and Salemi, F. and Samajdar, A. and Sammut, L. and Sampson, L. M. and Sanchez, E. J. and Sandberg, V. and Sandeen, B. and Sanders, G. H. and Sanders, J. R. and Sassolas, B. and Sathyaprakash, B. S. and Saulson, P. R. and Sauter, O. and Savage, R. L. and Sawadsky, A. and Schale, P. and Schilling, R. and Schmidt, J. and Schmidt, P. and Schnabel, R. and Schofield, R. M. S. and Sch\"onbeck, A. and Schreiber, E. and Schuette, D. and Schutz, B. F. and Scott, J. and Scott, S. M. and Sellers, D. and Sengupta, A. S. and Sentenac, D. and Sequino, V. and Sergeev, A. and Serna, G. and Setyawati, Y. and Sevigny, A. and Shaddock, D. A. and Shaffer, T. and Shah, S. and Shahriar, M. S. and Shaltev, M. and Shao, Z. and Shapiro, B. and Shawhan, P. and Sheperd, A. and Shoemaker, D. H. and Shoemaker, D. M. and Siellez, K. and Siemens, X. and Sigg, D. and Silva, A. D. and Simakov, D. and Singer, A. and Singer, L. P. and Singh, A. and Singh, R. and Singhal, A. and Sintes, A. M. and Slagmolen, B. J. J. and Smith, J. R. and Smith, M. R. and Smith, N. D. and Smith, R. J. E. and Son, E. J. and Sorazu, B. and Sorrentino, F. and Souradeep, T. and Srivastava, A. K. and Staley, A. and Steinke, M. and Steinlechner, J. and Steinlechner, S. and Steinmeyer, D. and Stephens, B. C. and Stevenson, S. P. and Stone, R. and Strain, K. A. and Straniero, N. and Stratta, G. and Strauss, N. A. and Strigin, S. and Sturani, R. and Stuver, A. L. and Summerscales, T. Z. and Sun, L. and Sutton, P. J. and Swinkels, B. L. and Szczepa\ifmmode \acute{n}\else \'{n}\fi{}czyk, M. J. and Tacca, M. and Talukder, D. and Tanner, D. B. and T\'apai, M. and Tarabrin, S. P. and Taracchini, A. and Taylor, R. and Theeg, T. and Thirugnanasambandam, M. P. and Thomas, E. G. and Thomas, M. and Thomas, P. and Thorne, K. A. and Thorne, K. S. and Thrane, E. and Tiwari, S. and Tiwari, V. and Tokmakov, K. V. and Tomlinson, C. and Tonelli, M. and Torres, C. V. and Torrie, C. I. and T\"oyr\"a, D. and Travasso, F. and Traylor, G. and Trifir\`o, D. and Tringali, M. C. and Trozzo, L. and Tse, M. and Turconi, M. and Tuyenbayev, D. and Ugolini, D. and Unnikrishnan, C. S. and Urban, A. L. and Usman, S. A. and Vahlbruch, H. and Vajente, G. and Valdes, G. and Vallisneri, M. and van Bakel, N. and van Beuzekom, M. and van den Brand, J. F. J. and Van Den Broeck, C. and Vander-Hyde, D. C. and van der Schaaf, L. and van Heijningen, J. V. and van Veggel, A. A. and Vardaro, M. and Vass, S. and Vas\'uth, M. and Vaulin, R. and Vecchio, A. and Vedovato, G. and Veitch, J. and Veitch, P. J. and Venkateswara, K. and Verkindt, D. and Vetrano, F. and Vicer\'e, A. and Vinciguerra, S. and Vine, D. J. and Vinet, J.-Y. and Vitale, S. and Vo, T. and Vocca, H. and Vorvick, C. and Voss, D. and Vousden, W. D. and Vyatchanin, S. P. and Wade, A. R. and Wade, L. E. and Wade, M. and Waldman, S. J. and Walker, M. and Wallace, L. and Walsh, S. and Wang, G. and Wang, H. and Wang, M. and Wang, X. and Wang, Y. and Ward, H. and Ward, R. L. and Warner, J. and Was, M. and Weaver, B. and Wei, L.-W. and Weinert, M. and Weinstein, A. J. and Weiss, R. and Welborn, T. and Wen, L. and We\ss{}els, P. and Westphal, T. and Wette, K. and Whelan, J. T. and Whitcomb, S. E. and White, D. J. and Whiting, B. F. and Wiesner, K. and Wilkinson, C. and Willems, P. A. and Williams, L. and Williams, R. D. and Williamson, A. R. and Willis, J. L. and Willke, B. and Wimmer, M. H. and Winkelmann, L. and Winkler, W. and Wipf, C. C. and Wiseman, A. G. and Wittel, H. and Woan, G. and Worden, J. and Wright, J. L. and Wu, G. and Yablon, J. and Yakushin, I. and Yam, W. and Yamamoto, H. and Yancey, C. C. and Yap, M. J. and Yu, H. and Yvert, M. and Zadro\ifmmode \dot{z}\else \.{z}\fi{}ny, A. and Zangrando, L. and Zanolin, M. and Zendri, J.-P. and Zevin, M. and Zhang, F. and Zhang, L. and Zhang, M. and Zhang, Y. and Zhao, C. and Zhou, M. and Zhou, Z. and Zhu, X. J. and Zucker, M. E. and Zuraw, S. E. and Zweizig, J.},
  collaboration = {LIGO Scientific Collaboration and Virgo Collaboration},
  journal = {Phys. Rev. Lett.},
  volume = {116},
  issue = {6},
  pages = {061102},
  numpages = {16},
  year = {2016},
  month = {Feb},
  publisher = {American Physical Society},
  doi = {10.1103/PhysRevLett.116.061102},
  url = {https://link.aps.org/doi/10.1103/PhysRevLett.116.061102}
}

@article{PhysRevX.13.021019,
  title = {Gravitational-Wave Detector for Postmerger Neutron Stars: Beyond the Quantum Loss Limit of the Fabry-Perot-Michelson Interferometer},
  author = {Zhang, Teng and Yang, Huan and Martynov, Denis and Schmidt, Patricia and Miao, Haixing},
  journal = {Phys. Rev. X},
  volume = {13},
  issue = {2},
  pages = {021019},
  numpages = {13},
  year = {2023},
  month = {May},
  publisher = {American Physical Society},
  doi = {10.1103/PhysRevX.13.021019},
  url = {https://link.aps.org/doi/10.1103/PhysRevX.13.021019}
}

@article{Black2001AnIT,
  title={An introduction to Pound–Drever–Hall laser frequency stabilization},
  author={Eric D. Black},
  journal={American Journal of Physics},
  year={2001},
  volume={69},
  pages={79-87},
  url={https://api.semanticscholar.org/CorpusID:45532388}
}

@article{Guo_2023,
doi = {10.1088/1361-6382/ad0454},
url = {https://doi.org/10.1088/1361-6382/ad0454},
year = {2023},
month = {oct},
publisher = {IOP Publishing},
volume = {40},
number = {23},
pages = {235005},
author = {Guo, Xinyao and Zhang, Teng and Martynov, Denis and Haixing, Miao},
title = {Sensing and control scheme for the inteferometer configuration with an L-shaped resonator},
journal = {Classical and Quantum Gravity}
}
\appendix
\section{Derivation of each readout}\label{appx:readout}
Here we provide the derivation of Eqs.~\ref{eq:rxy and ryx when re=1}--\ref{eq:intra-cavityDC when re<1}. With the notation in Fig.~\ref{fig:layoutall}, the reflected field of the beam propagating through the clockwise path and returning along the same path, namely portion \ding{174} in the figure, is expressed as \(r_{xx}E_\mathrm{in}\), where \(E_\mathrm{in}\) denotes the incident field and \(r_{xx}\) is 
the corresponding reflectivity. The first subscript indicates the incident path, and the second represents the returning path. Similarly, \(r_{yy}\) corresponds to the reflection of the beam impinging 
and returning along the counterclockwise path. 

Hence, the effective reflectivities of the L-shaped cavity for the two incident paths can be written as
\begin{equation}
r_{xx}=-\frac{t_i^2r_ee^{i\frac{2\omega L_x}{c}}}{1-r_i^2r_e^2e^{i2\phi}}, \quad
r_{yy}=-\frac{t_i^2r_ee^{i\frac{2\omega L_y}{c}}}{1-r_i^2r_e^2e^{i2\phi}},
\label{eq1}
\end{equation}
and
\begin{equation}
r_{yx}=r_{xy}=r_i-\frac{t_i^2r_ir_e^2e^{i2\phi}}{1-r_i^2r_e^2e^{i2\phi}}
=\frac{r_i-r_ir_e^2e^{i2\phi}}{1-r_i^2r_e^2e^{i2\phi}}.
\label{eq2}
\end{equation}

Here \(r_i\) and \(r_e\) denote the reflectivities of the input and end test masses (ITM and ETM) of the L-shaped cavity, respectively. The total phase is defined as \(\phi=\omega L_+/c\), where \(L_+\) is the L-shaped cavity common mode. The frequency \(\omega\) may correspond to either the carrier \(\omega_0\) or \(\omega_0\pm\omega_m\), where 
\(\omega_m\) is the modulation frequency of the radio-frequency (RF) sidebands, since both the 
carrier and the RF sidebands share the same functional form of effective reflectivity. 

For carrier, when \(L_+\) is locked at resonance, there is \(e^{i\phi}=1\), and the subsequent analysis is based on this con-
dition. In the ideal case where \(r_e=1\), Eqs.~(\ref{eq1})--(\ref{eq2}) yield 
\(r_{xy}=r_{yx}=0\), meaning that light entering from the \(x\) path returns entirely through the same path, and likewise for the \(y\) path. This is the derivation of Eqs.~\ref{eq:rxy and ryx when re=1} and \ref{eq:rxx and ryy when re=1}. The interferometer therefore becomes equivalent to a folded Michelson interferometer when viewed from outside the L-shaped cavity. When \(r_e<1\), however, \(r_{xy}\) and \(r_{yx}\) are nonzero; these components are referred to as the Sagnac mode, and \(r_{xy}=r_{yx}\) always holds. 

Consequently, the overall reflectivity of the interferometer for light input and 
output from the bright port is given by
\begin{equation}
\begin{aligned}
r_{br}
&=\tfrac12( e^{i2\psi_1}r_{xx}+e^{i(\psi_1+\psi_2)}r_{xy}
+e^{i(\psi_1+\psi_2)}r_{yx}+e^{i2\psi_2}r_{yy})\\
&=-\frac{t_i^2r_ee^{i\phi}}{1-r_i^2r_e^2e^{i2\phi}}\cos{\!\left(\frac{\omega\mathcal{L}_-}{c}\right)}
+\frac{r_i-r_ir_e^2e^{i2\phi}}{1-r_i^2r_e^2e^{i2\phi}},
\end{aligned}
\end{equation}
and the corresponding transmission to the dark port is
\begin{equation}
\begin{aligned}
t_{dr}
&=\tfrac12( e^{i2\psi_1}r_{xx}-e^{i(\psi_1+\psi_2)}r_{xy}
+e^{i(\psi_1+\psi_2)}r_{yx}-e^{i2\psi_2}r_{yy})\\
&=i2 e^{i(\psi_1+\psi_2)}
\frac{t_i^2r_e e^{i\phi}}{1-r_i^2r_e^2e^{i2\phi}}
\sin\!\left(\frac{\omega\mathcal{L}_-}{c}\right),
\end{aligned}
\end{equation}
where \(\mathcal{L}_-=L_-+l_-\), \(\psi_1=\omega l_x/c \), \(\psi_2=\omega l_y/c \) and \(\omega\) applies to both the carrier and the RF sidebands. 

Even in the presence of the Sagnac mode, the DC signal at the dark port remains a purely sinusoidal function, which is the characteristic of a Michelson-like interferometer. The DC signal at the dark port is therefore expressed as
\begin{equation}
\begin{aligned}
\frac{S_\mathrm{dark}}{S_\mathrm{in}}
&=J_0^2|t_{dr,c}|^2+J_1^2(|t_{dr,s+}|^2+|t_{dr,s-}|^2)\\
&\approx\frac{t_i^4r_e^2}{(1-r_i^2r_e^2)^2}
\sin^2\!\left(\frac{\omega_0 \mathcal{L}_-}{c}\right),
\end{aligned}
\end{equation}
where \(S_\mathrm{in}\) is the input optical power, and \(t_{dr,c}\), \(t_{dr,s+}\), and 
\(t_{dr,s-}\) are the transmissions of the carrier and the upper and lower sidebands, respectively. In our experiment, the modulation depth is quite small such that $J_1\ll J_0$. Consequently, when considering the DC signal, the sideband contribution can be safely neglected, which leads to Eq.~\ref{eq:darkDC when re<1}. 

Similarly, the DC signal at the bright port is given by
\begin{equation}
\begin{aligned}
\frac{S_\mathrm{bright}}{S_\mathrm{in}}
&=J_0^2|r_{br,c}|^2
+J_1^2(|r_{br,s+}|^2+|r_{br,s-}|^2)\\
&\approx
\left|-\frac{t_i^2r_e}{1-r_i^2r_e^2}
\cos\!\left(\frac{\omega_0\mathcal{L}_-}{c}\right)
+\frac{r_i-r_ir_e^2}{1-r_i^2r_e^2}\right|^2,
\end{aligned}
\end{equation}
which corresponds to Eq.~\ref{eq:brightDC when re<1}. Expanding the expression yields
\begin{equation}
\begin{aligned}
\frac{S_\mathrm{bright}}{S_\mathrm{in}}
=&\frac{t_i^4r_e^2}{(1-r_i^2r_e^2)^2}
\cos^2\!\left(\frac{\omega_0 \mathcal{L}_-}{c}\right)\\
&-2\frac{t_i^2t_e^2r_ir_e}{(1-r_i^2r_e^2)^2}
\cos\!\left(\frac{\omega_0 \mathcal{L}_-}{c}\right)
+\frac{r_i^2t_e^4}{(1-r_i^2r_e^2)^2}.
\end{aligned}
\end{equation}
It can be seen that the bright-port signal contains both first- and second-order cosine terms; 
therefore, the peak spacing corresponds to \(\lambda/2\), while the amplitude modulation has a period of \(\lambda\). 

Following the same procedure, the transmission coefficients from outside into the L-shaped cavity through the ITM
can be written as
\begin{equation}
\begin{aligned}
t_{xx}=\frac{t_ie^{i\frac{\omega L_x}{c}}}{1-r_i^2r_e^2e^{i2\phi}},\quad
t_{yx}=\frac{t_ie^{i\frac{\omega L_y}{c}}r_ir_e e^{i\phi}}{1-r_i^2r_e^2e^{i2\phi}}.
\end{aligned}
\end{equation}
$t_{xx}$ represents the transmission of light incident from the $x$ path into the $x$ arm of the L-shaped cavity, while $t_{yx}$ represents the transmission of light incident from the $y$ path into the $x$ arm of the L-shaped cavity. Besides, the sidebands are off-resonance in the cavity.
 Thus, the total power in the L-shaped cavity is given by
\begin{equation}
\begin{aligned}
\frac{S_\mathrm{cavity}}{S_\mathrm{in}}
&=\tfrac{1}{2}\left|e^{i\frac{\omega_0 l_x}{c}}t_{xx}+e^{i\frac{\omega_0 l_y}{c}}t_{yx}\right|^2\\
&=\frac{t_i^2}{2(1-r_i^2r_e^2)^2}
\!\left(1+r_i^2r_e^2+2r_ir_e
\cos{\!\left(\frac{\omega_0 \mathcal{L}_-}{c}\right)}\right),
\end{aligned}
\end{equation}
which corresponds to Eq.~\ref{eq:intra-cavityDC when re<1}. In experiment, we extract the intra-cavity power from one of the ETM transmission port, whose power is $S_\mathrm{trans}=t_e^2\,S_\mathrm{cavity}$. Since, for $r_e=1$, Eqs.~\ref{eq:darkDC when re<1}-\ref{eq:intra-cavityDC when re<1} degenerate to Eqs.~\ref{eq:darkDC and brightDC when re=1}-\ref{eq:intra-cavityDC when re=1}, no additional proof is required.
It is also straightforward to verify that if both ETMs are lossless (namely \(r_e^2+t_e^2=1\)), the relationship \(S_\mathrm{bright}+S_\mathrm{dark}+2S_\mathrm{trans}=1\) holds, which satisfies energy conservation and confirms the validity of the 
theoretical model.


\end{document}